\documentclass[twocolumn,preprintnumbers,amsmath,amssymb]{revtex4}

\usepackage{graphicx}

\usepackage{dcolumn}

\usepackage{bm}

\usepackage{multirow}

\usepackage{color}

\usepackage{amsmath}

\begin{document}

\title{Optical orientation of electron spins and valence band spectroscopy in germanium}
\author{F. Pezzoli}
\altaffiliation{fabio.pezzoli@unimib.it}
\affiliation{LNESS and Dipartimento di Scienza dei Materiali, Universit\`{a} degli Studi di Milano-Bicocca, via R. Cozzi 55, I-20125 Milano, Italy}

\author{A. Balocchi}
\affiliation{Universit\'{e} de Toulouse, INSA-CNRS-UPS, LPCNO, 135 Ave. de Rangueil, 31077 Toulouse, France}

\author{E. Vitiello}
\affiliation{LNESS and Dipartimento di Scienza dei Materiali, Universit\`{a} degli Studi di Milano-Bicocca, via R. Cozzi 55, I-20125 Milano, Italy}

\author{T. Amand}
\affiliation{Universit\'{e} de Toulouse, INSA-CNRS-UPS, LPCNO, 135 Ave. de Rangueil, 31077 Toulouse, France}

\author{X. Marie}
\affiliation{Universit\'{e} de Toulouse, INSA-CNRS-UPS, LPCNO, 135 Ave. de Rangueil, 31077 Toulouse, France}

\begin{abstract}
We have investigated optical orientation in the vicinity of the direct gap of bulk germanium. The electron spin polarization is studied via polarization-resolved photoluminescence excitation spectroscopy unfolding the interplay between doping and ultrafast electron transfer from the center of the Brillouin zone towards its edge. As a result, the direct-gap photoluminescence circular polarisation can vary from 30\% to -60\% when the excitation laser energy increases. This study provides also simultaneous access to the resonant electronic Raman scattering due to inter-valence band excitations of spin-polarized holes, yielding a fast and versatile spectroscopic approach for the determination of the energy spectrum of holes in semiconducting materials.
\end{abstract}


\maketitle

The dynamics of nonequilibrium spin polarized carriers has been extensively investigated in semiconductors with strong light-matter interaction \cite{Zutic04, dyakonov_spin_2008}. Prompt access to the spin-related phenomena in materials having a direct energy gap such as III-V and II-VI zinc-blend compounds has been initially achieved by means of optical orientation that is the transfer of angular momentum from the absorbed light to the photogenerated carriers \cite{Zutic04, Parsons69, Amand97, Dyakonov_OO, Urbaszek13}.

In indirect gap semiconductors, on the contrary, the strong electron-phonon interaction and the relatively long carrier lifetime hindered optical investigations and puzzled the physical picture of the preferential alignment and relaxation of the spin angular momentum \cite{Zutic04,Lampel68, Jansen12}. This has effectively slowed down the progress in the understanding of the spin physics in group IV elements, which are pivotal for the ultimate monolithic implementation of spin quantum-bits in the solid state as they offer record-long spin lifetimes \cite{li12, Saeedi13, Kawakami14, Veldhorst14, Muhonen14} and can benefit from the vast infrastructures of the mainstream Si-based electronics \cite{Fodor06, Jansen12, Zwanenburg13}.

Recently it has been pointed out that nonequilibrium spin populations can be optically induced in germanium since the conduction band (CB) minimum at the center ($\Gamma$) of the Brillouin zone lies only 140 meV higher in energy than the fundamental edge at the $L$ point \cite{Loren09, Guite11, Pezzoli12}. Owning to the multivalley nature of the CB of Ge, however, electrons photogenerated at $\Gamma$ experience a phonon-induced scattering towards the side valleys, corresponding to an effective relaxation process which leads to unexplored and intriguing consequences on the spin-related phenomena \cite{Bottegoni13, Rinaldi14, Giorgioni14}. The detailed understanding of the dynamics of spin polarized carriers and its entanglement with the energy relaxation mechanisms is indeed crucial and it can eventually unfold the potential of Ge within the spintronics arena \cite{Rioux10}. Notably very little is known about spin flip scattering by dopants \cite{Song14} and the role played by impurities in determining the emission of circularly polarized light from the optically coupled CB and valence band (VB) states \cite{Pezzoli13}. 

In this communication we address the optical orientation process via polarization-resolved photoluminescence  excitation (PLE) spectroscopy, providing insights into the electron spin dynamics in the multivalley CB of bulk Ge. At the direct gap absorption, an electron spin polarization as large as 50\% is measured, i.e. the maximum value in bulk materials. By injecting spin polarized electrons with a well-controlled excess energy at the zone center, we can infer how ultrafast electron-phonon scattering and doping-induced Coulomb interactions affect the spontaneous recombination mechanisms of electrons and holes across the direct edge ($E_0$ gap). We emphasize that previous photoluminescence (PL) measurements in bulk germanium were performed using a fixed laser excitation energy \cite{Pezzoli13, Giorgioni14}. The use of a tunable circularly polarized excitation allows furthermore the observation of resonant Electronic Raman Scattering (ERS), i.e. the inelastic light scattering by electronic excitations \cite{Olego80}. We provide the first report of scattering involving the split-off (SO) hole subband, and unambiguously demonstrate that inter-valence-band excitations are responsible for the radiative recombination of spin polarized carriers which do not experience relaxation in intermediate states. Finally, inspired by previous studies \cite{Wagner85, Tanaka94, Nazvanova00} we show that the measurement of the energy dependence of ERS in the vicinity of the $E_0$ gap yields an accurate spectroscopic tool for a complete mapping of the VB dispersion in semiconducting materials.

In order to unfold how optical spin orientation is affected either by the doping content or by the donor/acceptor nature of the guest element, we studied the following bulk Ge(001) wafers. We employed a $p$-type Ge sample, with a doping concentration of $3.6 \times 10^{18}$ $\mathrm{cm^{-3}}$, termed \textit{p}$^+ Ge$, and two Ge wafers with the same impurity content of $2.2 \times 10^{16}$ $\mathrm{cm}^{-3}$, but one $n$-type and the other $p$-type named \textit{n}$^- Ge$ and \textit{p}$^- Ge$, respectively. The doping levels have been obtained by means of room temperature resistivity measurements.

An optical parametric oscillator pumped by a mode-locked Ti:Sa laser was used as a tunable excitation source between about 1000 and 1400 nm. The laser worked at a repetition rate of  80 MHz with a pulse and a spectral width of about 1.4 ps and 1 meV, respectively. The laser spot size on the sample surface had a diameter of $\approx$100 $\mu$m, and the average excitation power has been kept constant at about 60 mW during all the experiments with the different samples. PL measurements were carried out at 6 K using a closed-cycle cryostat and a spectrometer equipped with a liquid-nitrogen-cooled InGaAs array detector, ensuring a spectral pitch $<0.4$ meV. The back scattered laser light was suppressed by long pass filters (LP). The luminescence polarization was probed by a quarter-wave plate followed by a polarizer placed in front of the spectrometer and for each photon energy of the exciting light, the intensity of the right- ($\sigma^+$) and left-handed ($\sigma^-$) circularly polarized PL was time integrated and provided the PL polarization $P_c = (I^+ - I^-) / (I^+ + I^-)$. Here $I^+$ and $I^-$ are the PL intensity components co- and counter-circularly polarized with respect to the laser. It is worth noting that a positive (negative) $P_c$ value corresponds to a PL that is copolarized (counterpolarized) with respect to the excitation. 

\begin{figure}
   \centering
   \includegraphics[width=8.6cm]{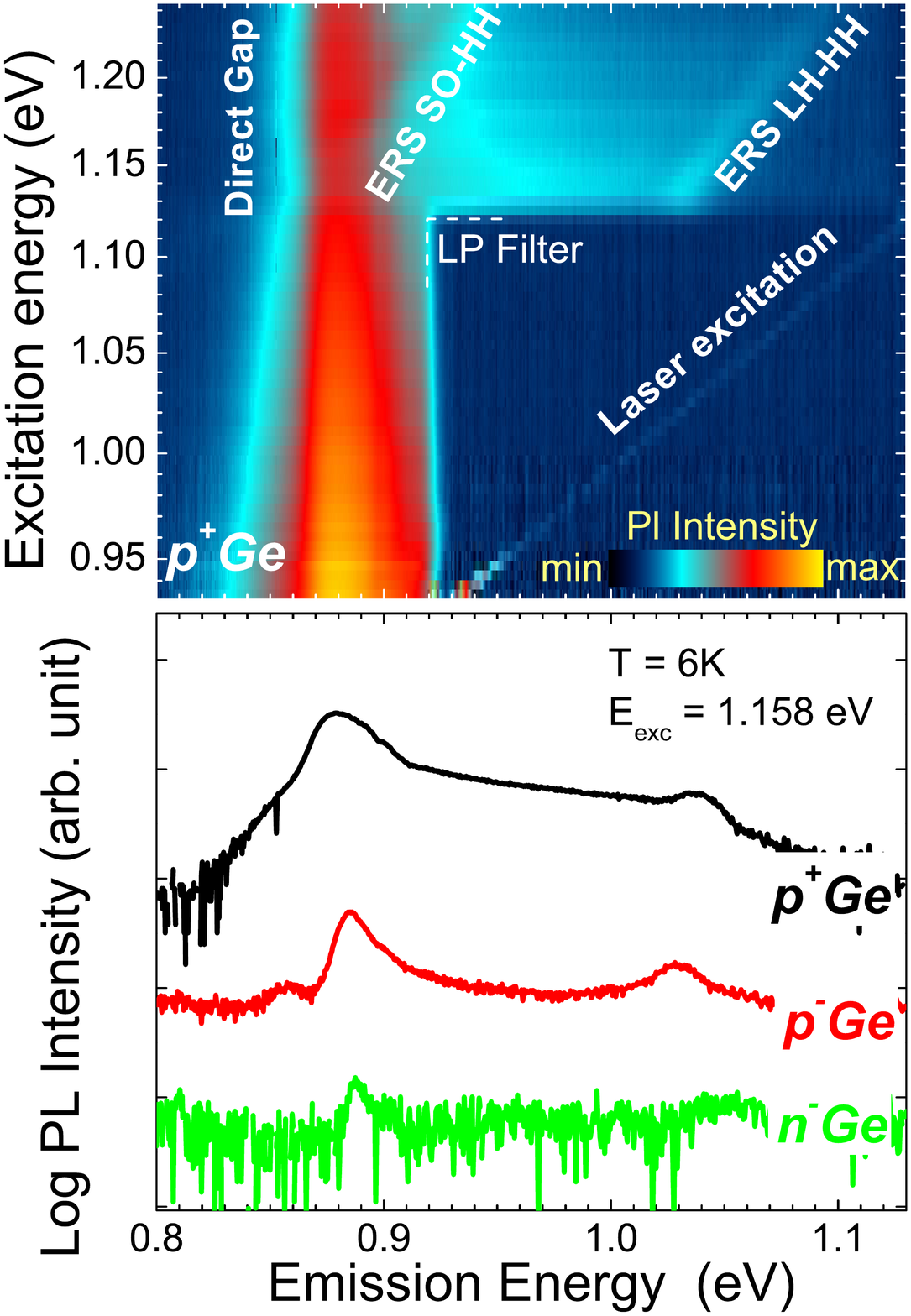}
   \caption{(color online). Upper panel: Contour plot of the PL intensity of \textit{p}$^+ Ge$ as a function of the excitation energy. The main direct gap peak and the inter-VB electronic Raman scattering (ERS) associated to heavy hole (HH) scattering to the light (LH) and splitoff (SO) subbands are shown along with the laser line and the cutoff due to a long pass filter (LP). Lower panel: Time integrated PL intensities for 
bulk Ge wafers having different doping, namely \textit{p}$^+ Ge$ $3.6 \times 10^{18}$ $\mathrm{cm^{-3}}$ (black curve), \textit{p}$^- Ge$ $2.2 \times 10^{16}$ $\mathrm{cm}^{-3}$ (red curve) and \textit{n}$^- Ge$ $2.2 \times 10^{16}$ $\mathrm{cm}^{-3}$ (green curve). The PL spectra were collected at 6K and for a cross-polarized emission with respect to the excitation at 1.158 eV. The spectra have been vertically shifted for clarity.} \label{fig:1}
\end{figure}

Figure~1 depicts the color coded map of the low temperature PLE of \textit{p}$^+ Ge$. The main and well-defined PL peak at 0.879 eV is the spontaneous radiative recombination across the direct gap $E_0$. This emission becomes weaker by increasing the excitation energy despite the larger joint density of states involved in the absorption process. Such finding can be interpreted as a result of the increased amount of electrons, which are photogenerated with a high excess energy in the CB and are more likely to be scattered out of the zone center \cite{Wagner84}. As it will be shown in the following, this has drastic consequences on the resulting spin population of $\Gamma$ valley electrons and holes and finally on the circular polarization degree of the $E_0$ PL. A closer look in Fig.~1 to the excitation range corresponding to the high-energy regime shows on the high energy side of the $E_0$ peak two additional bands, ERS HH-LH and ERS HH-SO. In contrast to $E_0$, these side bands shift to low energy by decreasing the incident photon energy. The ERS HH-SO eventually merges into the $E_0$ line and it contributes to the apparent broadening and change in lineshape of the direct gap emission. Noticeably, the spectrum reported in the lower panel of Fig.~1 for \textit{p}$^- Ge$, which possess a sharper $E_0$ line, suggests unexpectedly that a cross-over between $E_0$ and ERS HH-SO can take place with the latter transition occurring at energies smaller than the direct gap. Finally, Fig.~1 highlights that the ERS bands cannot be distinguished in the \textsl{n}-type material. All the aforementioned findings let us expect that ERS can be ascribed to light scattering phenomena due to electronic excitations occurring in the VB rather than to exciton-like band-to-band or band-to-impurity recombination.

\begin{figure*}
   \centering
   \includegraphics[angle=90,width=\textwidth]{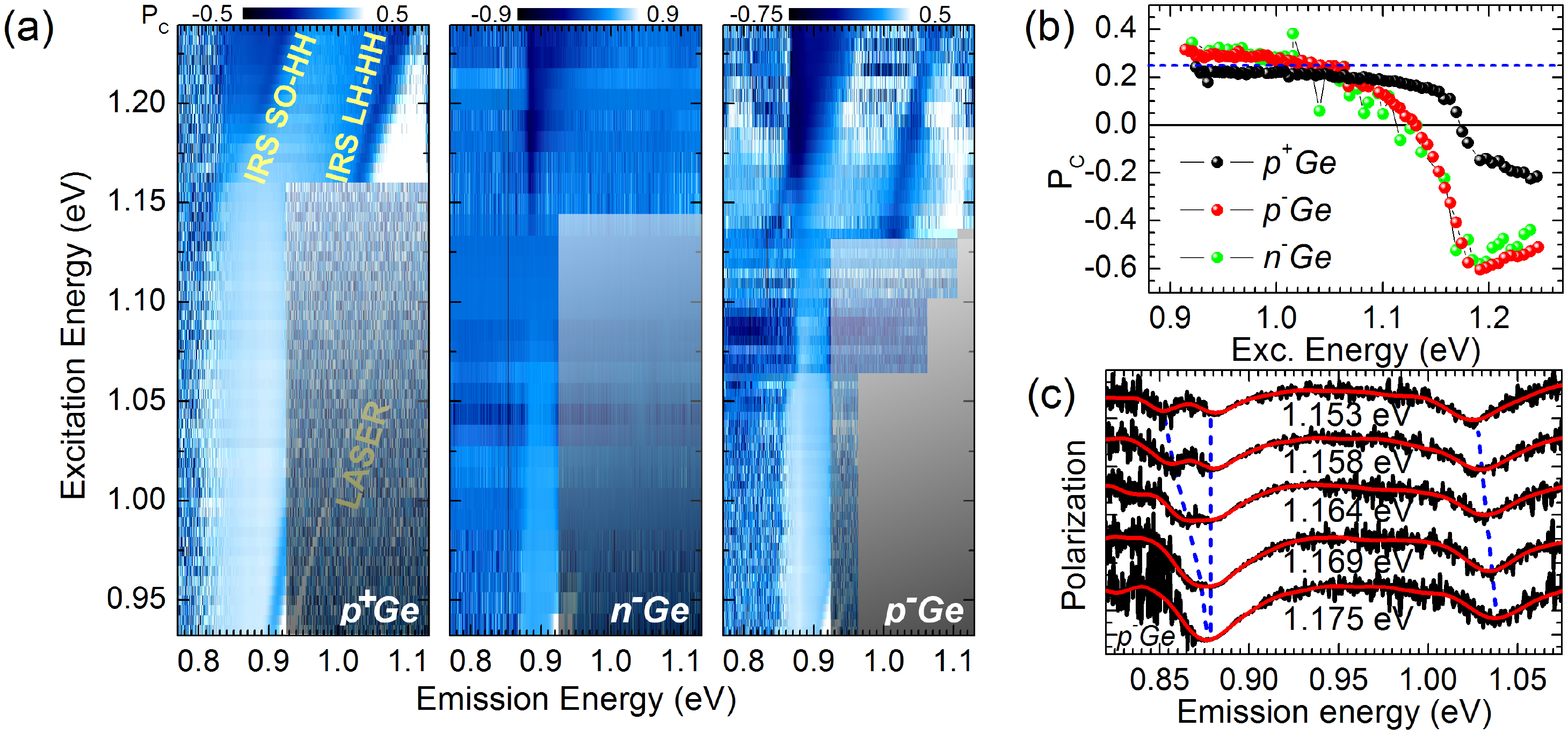}
   \caption{(color online). (a) Contour plot of the circular polarization degree $P_C$ of the PL of \textit{p}$^+ Ge$, \textit{n}$^- Ge$ and \textit{p}$^- Ge$ as a function of the excitation energy. The shadowed region of the map indicates data affected by the cutoff of a long pass filter. (b) Degree of circular polarization at the maximum of the luminescence $E_0$ peak as a function of the excitation energy. Data refers to a lattice temperature of 6K for \textit{p}$^+ Ge$ (orange dots), \textit{p}$^- Ge$ (blue dots) and \textit{n}$^- Ge$ (green dots). The blue dashed line defines the maximum $P_C = 0.25$ achievable in bulk material according to the zone center atomic-like character of the CB and VB states. (c) Close up of the $P_C$ data of \textit{p}$^- Ge$ in the 1.153-1.175 eV range of the excitation energy. The polarization spectra have been vertically shifted for clarity and the dotted blue lines are guide for the eye.} \label{fig:2}
\end{figure*}

The investigation of the state of polarization of the emitted light under circularly polarized excitation  contributes to clarify the dynamics of spin polarized carriers along with the origin of the observed spectral lines. 

We shall now concentrate on the direct gap transition. In the energy range spanned in Fig.~2(a) by the laser excitation, the $E_0$ PL turns out to be circularly polarized for all the studied samples. Fig.~2(b) further quantifies the values of $P_C$ evaluated at the maximum of the $E_0$ peak as a function of the exciting photon energy. $P_C$ reaches 0.22 in \textit{p}$^+ Ge$ at the threshold of direct gap absorption ($\approx 0.9$ eV). This is in good agreement with the $0.22\leq P_C \leq 0.25$ expected in this energy range by $\mathbf{k \cdot p}$ calculations of the electron spin polarization \cite{Rioux10}, and it is consistent with the fact that in bulk material the electron spin lifetime is typically longer than the hole spin relaxation time \cite{Dyakonov_OO, Giorgioni14}. We notice, however, that in the excitation energy range below 1.05 eV for \textit{p}$^- Ge$ and \textit{n}$^- Ge$, $P_C$ is systematically larger than in \textit{p}$^+ Ge$, being as high as 0.3. 

Furthermore by increasing the excitation energy the polarization degree of the $E_0$ band decreases in all the studied samples. In contrast to the theory which predicts in our energy window a positive electron spin polarization as low as 10\% \cite{Rioux10}, we found experimentally that $P_C$ changes sign between 1.1 and 1.2 eV becoming negative, that is cross-polarized with respect to the laser. The data of \textsl{p}-type wafers reported in Fig.~2(b) demonstrate that such helicity cross-over shifts to higher energies by increasing the doping content, while for samples with the same doping content, i.e. \textit{n}$^- Ge$ and \textit{p}$^- Ge$, it occurs at the same energy. In addition, the very similar degree of PL polarization observed in \textit{n}$^- Ge$ and \textit{p}$^- Ge$ over the whole excitation range discloses that the spin populations are remarkably unaffected by the donor or acceptor nature of the impurities. It should be noted that in these two samples the time-integrated circular polarization $P_C$ reaches -60\% at about 1.2 eV. This represents the largest value reported to date for band-to-band recombination in bulk zinc-blend semiconductors without relying on deformation potential or quantum confinement effects.

Such findings can be rationalized within the framework of the following physical picture \cite{Pezzoli13}. For a given energy and helicity of the absorbed photons, electrons will be promoted to the CB from the top of the VB or from the SO band with opposite spin orientations and with different excess energy with respect to the  bottom of the $\Gamma$ valley. Below 1.1 eV only electrons from optically coupled Heavy (HH) and Light (LH) Hole subbands contribute to the radiative recombination, yielding, in light of the dipole allowed selection rules \cite{Dyakonov_OO, Zutic04}, a $E_0$ PL that is copolarized with the excitation. As the incident photon energy is increased above 1.1 eV, vertical transitions will promote electrons at the CB edge starting from pristine SO levels that lie 290 meV below the top of the VB (see inset of Fig.~3). These electrons can then recombine radiatively giving rise to a counterpolarized PL component. At the same time, electrons originating from the top of the VB are excited into higher states in the CB and thus follow the more effective energy relaxation channel offered by the phonon-assisted intervalley scattering, which drives them out of the zone center and reduces the radiative recombination events that result in the copolarized $E_0$ PL \cite{Pezzoli13}. The dynamics of spin-polarized carriers discussed above is indeed responsible, along with the change of the $E_0$ intensity reported in Fig.~1, also for the continuous decrease of $P_C$ with the exciting photon energy. In particular, the helicity inversion shown in Fig.~2 is the fingerprint of the switched dominance within the spin ensemble at $\Gamma$ of the subset of electrons optically coupled to the SO (negative $P_C$) over the electrons coupled to the HH (positive $P_C$) subband. 

Impurities can activate additional mechanisms enriching the spin-related phenomena that are experimentally accessible. Indeed during the thermalization process, the Coulomb interactions of hot CB electrons with background carriers introduced by doping are known to enhance the probability of backward scattering from the satellite $X$ valleys towards the $\Gamma$ valley \cite{Mak94, Pezzoli13}. Such mechanism might play an important role since a larger impurity content can increase the weight of the copolarized contribution to the $E_0$ transitions. Consistently, our data  demonstrate that for the heavily doped \textit{p}$^+ Ge$ (i) the positive copolarized regime is stretched over a larger energy range than for the less-doped counterpart, i.e. \textit{p}$^- Ge$, eventually leading  to an helicity cross-over that takes place at higher energies and (ii) in the negative counterpolarized regime the absolute values of $P_C$ are smaller than in \textit{p}$^- Ge$. Since the cooling time for electrons in the satellite $X$ valleys takes place on a ps time scale \cite{Zhou94, Mak94}, whereas the spin relaxation time of holes is expected to be in the sub-ps time range \cite{Loren11}, electrons that experience backward scattering to the zone center are likely to recombine radiatively with unpolarized holes. Although this scenario applies well to \textit{p}$^+ Ge$, for the less doped \textit{p}$^- Ge$ and \textit{n}$^- Ge$, the fraction of electrons that underwent $\Gamma$-$X$-$\Gamma$ scattering is greatly reduced because of the less efficient role of impurities. Under this condition, as the excitation energy approaches the direct gap threshold, the average electron lifetime at $\Gamma$ is shortened and if comparable to the hole spin lifetime it can eventually lead to a PL polarization degree larger than the maximum of 0.25 expected from the theory \cite{Dyakonov_OO, Zutic04, Rioux10}. Our data shown in Fig.~2 support this interpretation since below 1.05 eV  $P_C\approx 0.25$ for \textit{p}$^+ Ge$ but $P_C = 0.3$ for \textit{p}$^- Ge$ and \textit{n}$^- Ge$. 

\begin{figure}
   \centering
   \includegraphics[width=8.6cm]{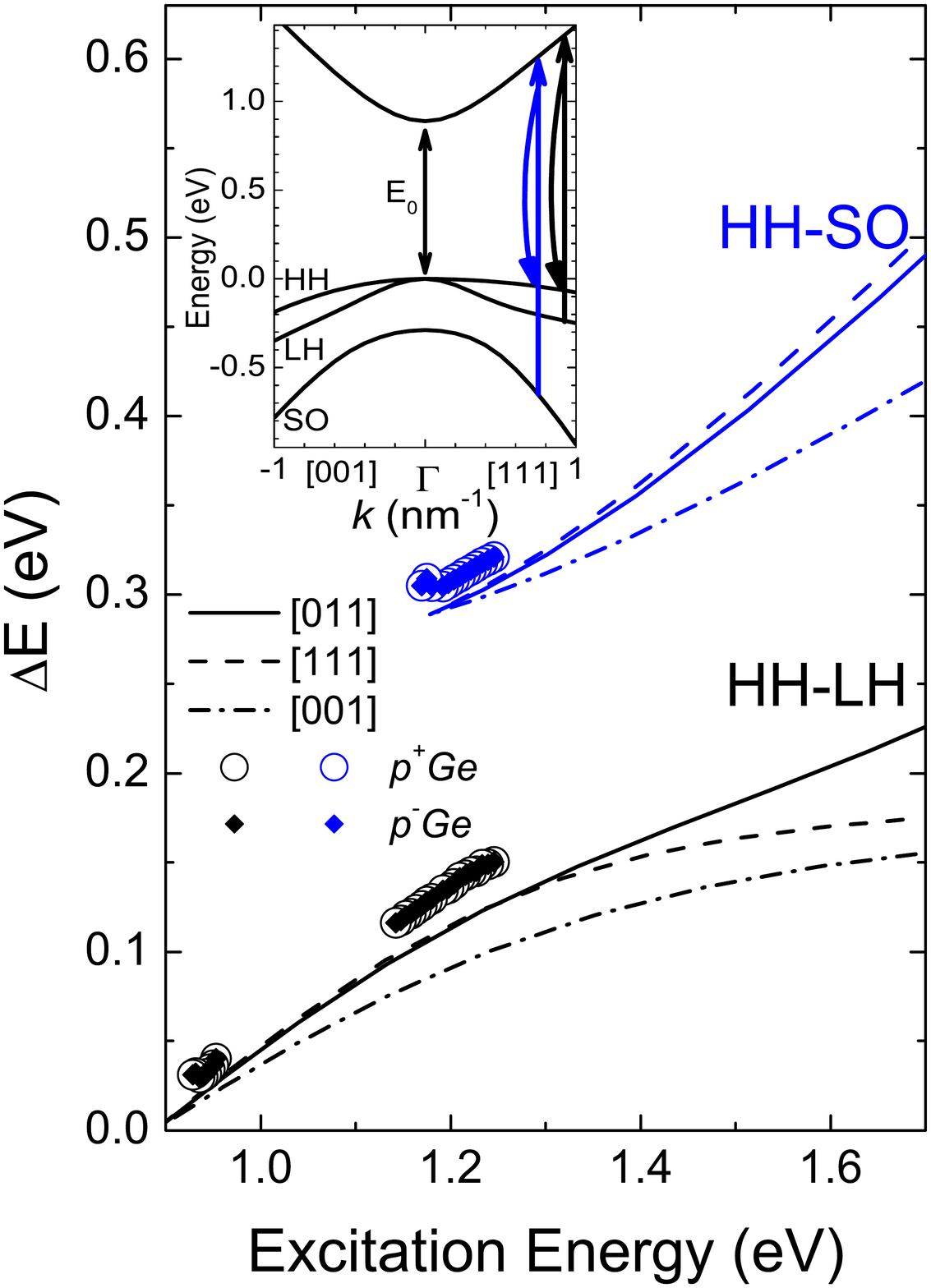}
   \caption{(color online). Energy difference $\Delta E$ between the ERS peaks and the laser energy. Data for \textit{p}$^+ Ge$ and \textit{p}$^- Ge$ are reported as a function of the incident photon energy as open circles and solid squares, respectively. The error bars are smaller than the size of the symbols. The lines corresponds to the calculated Raman peak positions due to HH-to-LH and HH-to-SO scattering along the $\left[001\right]$ (dashed-dotted line), $\left[011\right]$ (solid line) and $\left[111\right]$ (dashed line) directions in the momentum space. Wave vector-conserving transitions were assumed and reported as a functions of the laser energy. The inset shows an overview of the corresponding $\mathbf{k \cdot p}$ band structure of bulk Ge obtained in the vicinity of the zone center.} \label{fig:3}
\end{figure}

After having discussed the intertwined relationship between optical orientation, direct gap recombination and dynamics of spin polarized carrier, we shall focus on the mechanisms yielding the ERS bands observed in Fig.~1. According to the following arguments they can be ascribed to resonant inter-VB electronic Raman scattering. 

The data of \textit{p}$^- Ge$ reported in Fig.~2(a) and (c) favor this attribution since the ERS band at lower energy is washed out as soon as it sweeps across $E_0$. This is in line with a Raman process whose cross section gets greatly reduced for incident photon energies that are out-of-resonance with the electronic states of the crystal. The efficiency of the inter-VB Raman process is dictated by the free hole population \cite{Nazvanova00}, hence light scattering should be expected to be prominent in \textsl{p}-doped wafers. Indeed in Fig.~2(a), well-defined ERS features can be seen for \textit{p}$^+ Ge$ and \textit{p}$^- Ge$ and are absent when the majority charged carriers are electrons like in \textit{n}$^- Ge$. Moreover, the analysis of the dependence of $P_c$ upon the excitation wavelength, summarized in Fig.~2(a) and (c), demonstrates that the two ERS bands give rise to minima in the $P_C$ curves. The negative degree associated to the ERS peaks stems for photons whose angular momentum is the opposite of the one of the absorbed light, as expected for interband excitation of polarized holes in which the parity of the wave function of the positive carriers is preserved \cite{Uenoyama90}. For any incident photon energy, such HH-to-SO scattering is expected to lead to a spectral feature at a lower energy than the one due to HH-to-LH excitation because of the energy separation between the three hole subbands. As schematically shown in the inset of Fig.~3, these processes can be viewed as a spin polarized electron promoted out of a filled state of the SO (LH) subband into an intermediate CB state,  followed by a deexcitation onto an empty state in the heavy hole (HH) subband, while leaving in place the photoexcited hole in the original SO (LH) subband.  It should be noted that in this process the electron does not experience any energy relaxation in the CB and it can be safely assumed that the observed light scattering arises mainly from wave vector conserving transitions \cite{Tanaka94, Nazvanova00}. We point out that the well-satisfied polarization selection rules and the opposite sign with respect to the excitation suggest that the likely origin of the two lines is inter-VB excitation instead of hot luminescence \cite{Olego80}. 

We emphasize that the spectral position of the ERS peaks harbors important information about the electronic structure of the material. In Fig.~3 we summarized the measured values of the energy gap, $\Delta E$, between the laser and the two ERS peaks by varying the excitation photon energy. To gain further insights we calculated the band structure of bulk Ge along different directions in the crystal momentum space by means of 8-band $\mathbf{k \cdot p}$ perturbation theory (see Fig.~3). The resulting energy splitting, $\Delta E$, between the hole subbands is a function of Bloch's wave vector that in a wave vector-conserving Raman process can be recast into the laser energy that connects the initial state in the VB to the intermediate state in the CB by means of a vertical transition. The very good agreement between the experimental data and the calculations outlined in Fig.~3 further corroborates our explanation of the ERS bands, and it demonstrates that the full energy spectrum of holes can be straightforwardly derived from the optical measurement of the dependence of the electronic Raman lines upon the laser energy.
 
In conclusion, we used PLE spectroscopy to study how the dynamics of spin polarized electrons is perturbed by phonon-mediated intervalley scattering and Coulomb interactions with background carriers. We gathered access to inter-VB excitations mapping out the energy spectrum of holes, and we were able to clarify previous studies by providing the first evidence of the SO-to-HH scattering in bulk Ge. Looking ahead, our approach can be applied to shine light on how the multivalley structure of Ge can be modified by tin alloying to turn Ge into a direct-gap semiconductor. Finally, a more detailed understanding of the resonant electronic Raman scattering might provide insights into inter-VB absorption, recently pointed out as a possible bottleneck for optical gain in Ge. 

The authors acknowledge S. Mazzucato for technical assistance, A. Giorgioni for fruitful discussions and G. Isella for providing the low-doped samples. This work was supported by Fondazione Cariplo, grant 2013.0623. 


\begin{thebibliography}{33}
\expandafter\ifx\csname natexlab\endcsname\relax\def\natexlab#1{#1}\fi
\expandafter\ifx\csname bibnamefont\endcsname\relax
  \def\bibnamefont#1{#1}\fi
\expandafter\ifx\csname bibfnamefont\endcsname\relax
  \def\bibfnamefont#1{#1}\fi
\expandafter\ifx\csname citenamefont\endcsname\relax
  \def\citenamefont#1{#1}\fi
\expandafter\ifx\csname url\endcsname\relax
  \def\url#1{\texttt{#1}}\fi
\expandafter\ifx\csname urlprefix\endcsname\relax\def\urlprefix{URL }\fi
\providecommand{\bibinfo}[2]{#2}
\providecommand{\eprint}[2][]{\url{#2}}

\bibitem[{\citenamefont{Zutic et~al.}(2004)\citenamefont{Zutic, Fabian, and
  Das~Sarma}}]{Zutic04}
\bibinfo{author}{\bibfnamefont{I.}~\bibnamefont{Zutic}},
  \bibinfo{author}{\bibfnamefont{J.}~\bibnamefont{Fabian}}, \bibnamefont{and}
  \bibinfo{author}{\bibfnamefont{S.}~\bibnamefont{Das~Sarma}},
  \bibinfo{journal}{Rev. Mod. Phys.} \textbf{\bibinfo{volume}{76}},
  \bibinfo{pages}{323} (\bibinfo{year}{2004}).

\bibitem[{\citenamefont{Dyakonov}(2008)}]{dyakonov_spin_2008}
\bibinfo{editor}{\bibfnamefont{M.~I.} \bibnamefont{Dyakonov}}, ed.,
  \emph{\bibinfo{title}{Spin {Physics} in {Semiconductors}}}
  (\bibinfo{publisher}{Springer}, \bibinfo{address}{Berlin},
  \bibinfo{year}{2008}), \bibinfo{edition}{2008th} ed., ISBN
  \bibinfo{isbn}{9783540788195}.

\bibitem[{\citenamefont{Parsons}(1969)}]{Parsons69}
\bibinfo{author}{\bibfnamefont{R.~R.} \bibnamefont{Parsons}},
  \bibinfo{journal}{Phys. Rev. Lett.} \textbf{\bibinfo{volume}{23}},
  \bibinfo{pages}{1152} (\bibinfo{year}{1969}).

\bibitem[{\citenamefont{Amand et~al.}(1997)\citenamefont{Amand, Marie,
  Le~Jeune, Brousseau, Robart, Barrau, and Planel}}]{Amand97}
\bibinfo{author}{\bibfnamefont{T.}~\bibnamefont{Amand}},
  \bibinfo{author}{\bibfnamefont{X.}~\bibnamefont{Marie}},
  \bibinfo{author}{\bibfnamefont{P.}~\bibnamefont{Le~Jeune}},
  \bibinfo{author}{\bibfnamefont{M.}~\bibnamefont{Brousseau}},
  \bibinfo{author}{\bibfnamefont{D.}~\bibnamefont{Robart}},
  \bibinfo{author}{\bibfnamefont{J.}~\bibnamefont{Barrau}}, \bibnamefont{and}
  \bibinfo{author}{\bibfnamefont{R.}~\bibnamefont{Planel}},
  \bibinfo{journal}{Phys. Rev. Lett.} \textbf{\bibinfo{volume}{78}},
  \bibinfo{pages}{1355} (\bibinfo{year}{1997}).

\bibitem[{\citenamefont{Dyakonov and Perel}(1984)}]{Dyakonov_OO}
\bibinfo{author}{\bibfnamefont{M.~I.} \bibnamefont{Dyakonov}} \bibnamefont{and}
  \bibinfo{author}{\bibfnamefont{V.~I.} \bibnamefont{Perel}},
  \emph{\bibinfo{title}{Optical Orientation}} (\bibinfo{address}{North-Holland,
  New York}, \bibinfo{year}{1984}), chap.~\bibinfo{chapter}{2}.

\bibitem[{\citenamefont{Urbaszek et~al.}(2013)\citenamefont{Urbaszek, Marie,
  Amand, Krebs, Voisin, Maletinsky, H\"ogele, and Imamoglu}}]{Urbaszek13}
\bibinfo{author}{\bibfnamefont{B.}~\bibnamefont{Urbaszek}},
  \bibinfo{author}{\bibfnamefont{X.}~\bibnamefont{Marie}},
  \bibinfo{author}{\bibfnamefont{T.}~\bibnamefont{Amand}},
  \bibinfo{author}{\bibfnamefont{O.}~\bibnamefont{Krebs}},
  \bibinfo{author}{\bibfnamefont{P.}~\bibnamefont{Voisin}},
  \bibinfo{author}{\bibfnamefont{P.}~\bibnamefont{Maletinsky}},
  \bibinfo{author}{\bibfnamefont{A.}~\bibnamefont{H\"ogele}}, \bibnamefont{and}
  \bibinfo{author}{\bibfnamefont{A.}~\bibnamefont{Imamoglu}},
  \bibinfo{journal}{Rev. Mod. Phys.} \textbf{\bibinfo{volume}{85}},
  \bibinfo{pages}{79} (\bibinfo{year}{2013}).

\bibitem[{\citenamefont{Lampel}(1968)}]{Lampel68}
\bibinfo{author}{\bibfnamefont{G.}~\bibnamefont{Lampel}},
  \bibinfo{journal}{Phys. Rev. Lett.} \textbf{\bibinfo{volume}{20}},
  \bibinfo{pages}{491} (\bibinfo{year}{1968}).

\bibitem[{\citenamefont{Jansen}(2012)}]{Jansen12}
\bibinfo{author}{\bibfnamefont{R.}~\bibnamefont{Jansen}},
  \bibinfo{journal}{Nature Mater.} \textbf{\bibinfo{volume}{11}},
  \bibinfo{pages}{400} (\bibinfo{year}{2012}).

\bibitem[{\citenamefont{Li et~al.}(2012)\citenamefont{Li, Song, and
  Dery}}]{li12}
\bibinfo{author}{\bibfnamefont{P.}~\bibnamefont{Li}},
  \bibinfo{author}{\bibfnamefont{Y.}~\bibnamefont{Song}}, \bibnamefont{and}
  \bibinfo{author}{\bibfnamefont{H.}~\bibnamefont{Dery}},
  \bibinfo{journal}{Phys. Rev. B} \textbf{\bibinfo{volume}{86}},
  \bibinfo{pages}{085202} (\bibinfo{year}{2012}).

\bibitem[{\citenamefont{Saeedi et~al.}(2013)\citenamefont{Saeedi, Simmons,
  Salvail, Dluhy, Riemann, Abrosimov, Becker, Pohl, Morton, and
  Thewalt}}]{Saeedi13}
\bibinfo{author}{\bibfnamefont{K.}~\bibnamefont{Saeedi}},
  \bibinfo{author}{\bibfnamefont{S.}~\bibnamefont{Simmons}},
  \bibinfo{author}{\bibfnamefont{J.~Z.} \bibnamefont{Salvail}},
  \bibinfo{author}{\bibfnamefont{P.}~\bibnamefont{Dluhy}},
  \bibinfo{author}{\bibfnamefont{H.}~\bibnamefont{Riemann}},
  \bibinfo{author}{\bibfnamefont{N.~V.} \bibnamefont{Abrosimov}},
  \bibinfo{author}{\bibfnamefont{P.}~\bibnamefont{Becker}},
  \bibinfo{author}{\bibfnamefont{H.-J.} \bibnamefont{Pohl}},
  \bibinfo{author}{\bibfnamefont{J.~J.~L.} \bibnamefont{Morton}},
  \bibnamefont{and} \bibinfo{author}{\bibfnamefont{M.~L.~W.}
  \bibnamefont{Thewalt}}, \bibinfo{journal}{Science}
  \textbf{\bibinfo{volume}{342}}, \bibinfo{pages}{830} (\bibinfo{year}{2013}).

\bibitem[{\citenamefont{Kawakami et~al.}(2014)\citenamefont{Kawakami, Scarlino,
  Ward, Braakman, Savage, Lagally, Friesen, Coppersmith, Eriksson, and
  Vandersypen}}]{Kawakami14}
\bibinfo{author}{\bibfnamefont{E.}~\bibnamefont{Kawakami}},
  \bibinfo{author}{\bibfnamefont{P.}~\bibnamefont{Scarlino}},
  \bibinfo{author}{\bibfnamefont{D.~R.} \bibnamefont{Ward}},
  \bibinfo{author}{\bibfnamefont{F.~R.} \bibnamefont{Braakman}},
  \bibinfo{author}{\bibfnamefont{D.~E.} \bibnamefont{Savage}},
  \bibinfo{author}{\bibfnamefont{M.~G.} \bibnamefont{Lagally}},
  \bibinfo{author}{\bibfnamefont{M.}~\bibnamefont{Friesen}},
  \bibinfo{author}{\bibfnamefont{S.~N.} \bibnamefont{Coppersmith}},
  \bibinfo{author}{\bibfnamefont{M.~A.} \bibnamefont{Eriksson}},
  \bibnamefont{and} \bibinfo{author}{\bibfnamefont{L.~M.~K.}
  \bibnamefont{Vandersypen}}, \bibinfo{journal}{Nature Nanotech.}
  \textbf{\bibinfo{volume}{9}}, \bibinfo{pages}{666 } (\bibinfo{year}{2014}).

\bibitem[{\citenamefont{Veldhorst et~al.}(2014)\citenamefont{Veldhorst, Hwang,
  Yang, Leenstra, de~Ronde, Dehollain, Muhonen, Hudson, Itoh, Morello
  et~al.}}]{Veldhorst14}
\bibinfo{author}{\bibfnamefont{M.}~\bibnamefont{Veldhorst}},
  \bibinfo{author}{\bibfnamefont{J.~C.~C.} \bibnamefont{Hwang}},
  \bibinfo{author}{\bibfnamefont{C.~H.} \bibnamefont{Yang}},
  \bibinfo{author}{\bibfnamefont{A.~W.} \bibnamefont{Leenstra}},
  \bibinfo{author}{\bibfnamefont{B.}~\bibnamefont{de~Ronde}},
  \bibinfo{author}{\bibfnamefont{J.~P.} \bibnamefont{Dehollain}},
  \bibinfo{author}{\bibfnamefont{J.~T.} \bibnamefont{Muhonen}},
  \bibinfo{author}{\bibfnamefont{F.~E.} \bibnamefont{Hudson}},
  \bibinfo{author}{\bibfnamefont{K.~M.} \bibnamefont{Itoh}},
  \bibinfo{author}{\bibfnamefont{A.}~\bibnamefont{Morello}},
  \bibnamefont{et~al.}, \bibinfo{journal}{Nature Nanotech.}
  \textbf{\bibinfo{volume}{9}}, \bibinfo{pages}{981 } (\bibinfo{year}{2014}).

\bibitem[{\citenamefont{Muhonen et~al.}(2014)\citenamefont{Muhonen, Dehollain,
  Laucht, Hudson, Kalra, Sekiguchi, Itoh, Jamieson, McCallum, Dzurak
  et~al.}}]{Muhonen14}
\bibinfo{author}{\bibfnamefont{J.~T.} \bibnamefont{Muhonen}},
  \bibinfo{author}{\bibfnamefont{J.~P.} \bibnamefont{Dehollain}},
  \bibinfo{author}{\bibfnamefont{A.}~\bibnamefont{Laucht}},
  \bibinfo{author}{\bibfnamefont{F.~E.} \bibnamefont{Hudson}},
  \bibinfo{author}{\bibfnamefont{R.}~\bibnamefont{Kalra}},
  \bibinfo{author}{\bibfnamefont{T.}~\bibnamefont{Sekiguchi}},
  \bibinfo{author}{\bibfnamefont{K.~M.} \bibnamefont{Itoh}},
  \bibinfo{author}{\bibfnamefont{D.~N.} \bibnamefont{Jamieson}},
  \bibinfo{author}{\bibfnamefont{J.~C.} \bibnamefont{McCallum}},
  \bibinfo{author}{\bibfnamefont{A.~S.} \bibnamefont{Dzurak}},
  \bibnamefont{et~al.}, \bibinfo{journal}{Nature Nanotech.}
  \textbf{\bibinfo{volume}{9}}, \bibinfo{pages}{986 } (\bibinfo{year}{2014}).

\bibitem[{\citenamefont{Fodor and Levy}(2006)}]{Fodor06}
\bibinfo{author}{\bibfnamefont{P.~S.} \bibnamefont{Fodor}} \bibnamefont{and}
  \bibinfo{author}{\bibfnamefont{J.}~\bibnamefont{Levy}}, \bibinfo{journal}{J.
  Phys.-Condens. Mat.} \textbf{\bibinfo{volume}{18}}, \bibinfo{pages}{S745}
  (\bibinfo{year}{2006}).

\bibitem[{\citenamefont{Zwanenburg et~al.}(2013)\citenamefont{Zwanenburg,
  Dzurak, Morello, Simmons, Hollenberg, Klimeck, Rogge, Coppersmith, and
  Eriksson}}]{Zwanenburg13}
\bibinfo{author}{\bibfnamefont{F.~A.} \bibnamefont{Zwanenburg}},
  \bibinfo{author}{\bibfnamefont{A.~S.} \bibnamefont{Dzurak}},
  \bibinfo{author}{\bibfnamefont{A.}~\bibnamefont{Morello}},
  \bibinfo{author}{\bibfnamefont{M.~Y.} \bibnamefont{Simmons}},
  \bibinfo{author}{\bibfnamefont{L.~C.~L.} \bibnamefont{Hollenberg}},
  \bibinfo{author}{\bibfnamefont{G.}~\bibnamefont{Klimeck}},
  \bibinfo{author}{\bibfnamefont{S.}~\bibnamefont{Rogge}},
  \bibinfo{author}{\bibfnamefont{S.~N.} \bibnamefont{Coppersmith}},
  \bibnamefont{and} \bibinfo{author}{\bibfnamefont{M.~A.}
  \bibnamefont{Eriksson}}, \bibinfo{journal}{Rev. Mod. Phys.}
  \textbf{\bibinfo{volume}{85}}, \bibinfo{pages}{961} (\bibinfo{year}{2013}).

\bibitem[{\citenamefont{Loren et~al.}(2009)\citenamefont{Loren, Ruzicka,
  Werake, Zhao, van Driel, and Smirl}}]{Loren09}
\bibinfo{author}{\bibfnamefont{E.~J.} \bibnamefont{Loren}},
  \bibinfo{author}{\bibfnamefont{B.~A.} \bibnamefont{Ruzicka}},
  \bibinfo{author}{\bibfnamefont{L.~K.} \bibnamefont{Werake}},
  \bibinfo{author}{\bibfnamefont{H.}~\bibnamefont{Zhao}},
  \bibinfo{author}{\bibfnamefont{H.~M.} \bibnamefont{van Driel}},
  \bibnamefont{and} \bibinfo{author}{\bibfnamefont{A.~L.} \bibnamefont{Smirl}},
  \bibinfo{journal}{Appl. Phys. Lett.} \textbf{\bibinfo{volume}{95}},
  \bibinfo{pages}{092107} (\bibinfo{year}{2009}).

\bibitem[{\citenamefont{Guite and Venkataraman}(2011)}]{Guite11}
\bibinfo{author}{\bibfnamefont{C.}~\bibnamefont{Guite}} \bibnamefont{and}
  \bibinfo{author}{\bibfnamefont{V.}~\bibnamefont{Venkataraman}},
  \bibinfo{journal}{Phys. Rev. Lett.} \textbf{\bibinfo{volume}{107}},
  \bibinfo{pages}{166603} (\bibinfo{year}{2011}).

\bibitem[{\citenamefont{Pezzoli et~al.}(2012)\citenamefont{Pezzoli, Bottegoni,
  Trivedi, Ciccacci, Giorgioni, Li, Cecchi, Grilli, Song, Guzzi
  et~al.}}]{Pezzoli12}
\bibinfo{author}{\bibfnamefont{F.}~\bibnamefont{Pezzoli}},
  \bibinfo{author}{\bibfnamefont{F.}~\bibnamefont{Bottegoni}},
  \bibinfo{author}{\bibfnamefont{D.}~\bibnamefont{Trivedi}},
  \bibinfo{author}{\bibfnamefont{F.}~\bibnamefont{Ciccacci}},
  \bibinfo{author}{\bibfnamefont{A.}~\bibnamefont{Giorgioni}},
  \bibinfo{author}{\bibfnamefont{P.}~\bibnamefont{Li}},
  \bibinfo{author}{\bibfnamefont{S.}~\bibnamefont{Cecchi}},
  \bibinfo{author}{\bibfnamefont{E.}~\bibnamefont{Grilli}},
  \bibinfo{author}{\bibfnamefont{Y.}~\bibnamefont{Song}},
  \bibinfo{author}{\bibfnamefont{M.}~\bibnamefont{Guzzi}},
  \bibnamefont{et~al.}, \bibinfo{journal}{Phys. Rev. Lett.}
  \textbf{\bibinfo{volume}{108}}, \bibinfo{pages}{156603}
  (\bibinfo{year}{2012}).

\bibitem[{\citenamefont{Bottegoni et~al.}(2013)\citenamefont{Bottegoni,
  Ferrari, Cecchi, Finazzi, Ciccacci, and Isella}}]{Bottegoni13}
\bibinfo{author}{\bibfnamefont{F.}~\bibnamefont{Bottegoni}},
  \bibinfo{author}{\bibfnamefont{A.}~\bibnamefont{Ferrari}},
  \bibinfo{author}{\bibfnamefont{S.}~\bibnamefont{Cecchi}},
  \bibinfo{author}{\bibfnamefont{M.}~\bibnamefont{Finazzi}},
  \bibinfo{author}{\bibfnamefont{F.}~\bibnamefont{Ciccacci}}, \bibnamefont{and}
  \bibinfo{author}{\bibfnamefont{G.}~\bibnamefont{Isella}},
  \bibinfo{journal}{Appl. Phys. Lett.} \textbf{\bibinfo{volume}{102}},
  \bibinfo{eid}{152411} (\bibinfo{year}{2013}).

\bibitem[{\citenamefont{Rinaldi et~al.}(2014)\citenamefont{Rinaldi, Cantoni,
  Marangoni, Manzoni, Cerullo, and Bertacco}}]{Rinaldi14}
\bibinfo{author}{\bibfnamefont{C.}~\bibnamefont{Rinaldi}},
  \bibinfo{author}{\bibfnamefont{M.}~\bibnamefont{Cantoni}},
  \bibinfo{author}{\bibfnamefont{M.}~\bibnamefont{Marangoni}},
  \bibinfo{author}{\bibfnamefont{C.}~\bibnamefont{Manzoni}},
  \bibinfo{author}{\bibfnamefont{G.}~\bibnamefont{Cerullo}}, \bibnamefont{and}
  \bibinfo{author}{\bibfnamefont{R.}~\bibnamefont{Bertacco}},
  \bibinfo{journal}{Phys. Rev. B} \textbf{\bibinfo{volume}{90}},
  \bibinfo{pages}{161304} (\bibinfo{year}{2014}).

\bibitem[{\citenamefont{Giorgioni et~al.}(2014)\citenamefont{Giorgioni,
  Vitiello, Grilli, Guzzi, and Pezzoli}}]{Giorgioni14}
\bibinfo{author}{\bibfnamefont{A.}~\bibnamefont{Giorgioni}},
  \bibinfo{author}{\bibfnamefont{E.}~\bibnamefont{Vitiello}},
  \bibinfo{author}{\bibfnamefont{E.}~\bibnamefont{Grilli}},
  \bibinfo{author}{\bibfnamefont{M.}~\bibnamefont{Guzzi}}, \bibnamefont{and}
  \bibinfo{author}{\bibfnamefont{F.}~\bibnamefont{Pezzoli}},
  \bibinfo{journal}{Appl. Phys. Lett.} \textbf{\bibinfo{volume}{105}},
  \bibinfo{eid}{152404} (\bibinfo{year}{2014}).

\bibitem[{\citenamefont{Rioux and Sipe}(2010)}]{Rioux10}
\bibinfo{author}{\bibfnamefont{J.}~\bibnamefont{Rioux}} \bibnamefont{and}
  \bibinfo{author}{\bibfnamefont{J.~E.} \bibnamefont{Sipe}},
  \bibinfo{journal}{Phys. Rev. B} \textbf{\bibinfo{volume}{81}},
  \bibinfo{pages}{155215} (\bibinfo{year}{2010}).

\bibitem[{\citenamefont{Song et~al.}(2014)\citenamefont{Song, Chalaev, and
  Dery}}]{Song14}
\bibinfo{author}{\bibfnamefont{Y.}~\bibnamefont{Song}},
  \bibinfo{author}{\bibfnamefont{O.}~\bibnamefont{Chalaev}}, \bibnamefont{and}
  \bibinfo{author}{\bibfnamefont{H.}~\bibnamefont{Dery}},
  \bibinfo{journal}{Phys. Rev. Lett.} \textbf{\bibinfo{volume}{113}},
  \bibinfo{pages}{167201} (\bibinfo{year}{2014}).

\bibitem[{\citenamefont{Pezzoli et~al.}(2013)\citenamefont{Pezzoli, Qing,
  Giorgioni, Isella, Grilli, Guzzi, and Dery}}]{Pezzoli13}
\bibinfo{author}{\bibfnamefont{F.}~\bibnamefont{Pezzoli}},
  \bibinfo{author}{\bibfnamefont{L.}~\bibnamefont{Qing}},
  \bibinfo{author}{\bibfnamefont{A.}~\bibnamefont{Giorgioni}},
  \bibinfo{author}{\bibfnamefont{G.}~\bibnamefont{Isella}},
  \bibinfo{author}{\bibfnamefont{E.}~\bibnamefont{Grilli}},
  \bibinfo{author}{\bibfnamefont{M.}~\bibnamefont{Guzzi}}, \bibnamefont{and}
  \bibinfo{author}{\bibfnamefont{H.}~\bibnamefont{Dery}},
  \bibinfo{journal}{Phys. Rev. B} \textbf{\bibinfo{volume}{88}},
  \bibinfo{pages}{045204} (\bibinfo{year}{2013}).

\bibitem[{\citenamefont{Olego et~al.}(1980)\citenamefont{Olego, Cardona, and
  R\"ossler}}]{Olego80}
\bibinfo{author}{\bibfnamefont{D.}~\bibnamefont{Olego}},
  \bibinfo{author}{\bibfnamefont{M.}~\bibnamefont{Cardona}}, \bibnamefont{and}
  \bibinfo{author}{\bibfnamefont{U.}~\bibnamefont{R\"ossler}},
  \bibinfo{journal}{Phys. Rev. B} \textbf{\bibinfo{volume}{22}},
  \bibinfo{pages}{1905} (\bibinfo{year}{1980}).

\bibitem[{\citenamefont{Wagner and Cardona}(1985)}]{Wagner85}
\bibinfo{author}{\bibfnamefont{J.}~\bibnamefont{Wagner}} \bibnamefont{and}
  \bibinfo{author}{\bibfnamefont{M.}~\bibnamefont{Cardona}},
  \bibinfo{journal}{Phys. Rev. B} \textbf{\bibinfo{volume}{32}},
  \bibinfo{pages}{8071} (\bibinfo{year}{1985}).

\bibitem[{\citenamefont{Tanaka et~al.}(1994)\citenamefont{Tanaka, Ohtake, and
  Suemoto}}]{Tanaka94}
\bibinfo{author}{\bibfnamefont{K.}~\bibnamefont{Tanaka}},
  \bibinfo{author}{\bibfnamefont{H.}~\bibnamefont{Ohtake}}, \bibnamefont{and}
  \bibinfo{author}{\bibfnamefont{T.}~\bibnamefont{Suemoto}},
  \bibinfo{journal}{Phys. Rev. B} \textbf{\bibinfo{volume}{50}},
  \bibinfo{pages}{10694} (\bibinfo{year}{1994}).

\bibitem[{\citenamefont{Nazvanova et~al.}(2000)\citenamefont{Nazvanova,
  Suemoto, Maruyama, and Takano}}]{Nazvanova00}
\bibinfo{author}{\bibfnamefont{E.}~\bibnamefont{Nazvanova}},
  \bibinfo{author}{\bibfnamefont{T.}~\bibnamefont{Suemoto}},
  \bibinfo{author}{\bibfnamefont{S.}~\bibnamefont{Maruyama}}, \bibnamefont{and}
  \bibinfo{author}{\bibfnamefont{Y.}~\bibnamefont{Takano}},
  \bibinfo{journal}{Phys. Rev. B} \textbf{\bibinfo{volume}{62}},
  \bibinfo{pages}{1873} (\bibinfo{year}{2000}).

\bibitem[{\citenamefont{Wagner and Vina}(1984)}]{Wagner84}
\bibinfo{author}{\bibfnamefont{J.}~\bibnamefont{Wagner}} \bibnamefont{and}
  \bibinfo{author}{\bibfnamefont{L.}~\bibnamefont{Vina}},
  \bibinfo{journal}{Phys. Rev. B} \textbf{\bibinfo{volume}{30}},
  \bibinfo{pages}{7030} (\bibinfo{year}{1984}).

\bibitem[{\citenamefont{Mak and Vandriel}(1994)}]{Mak94}
\bibinfo{author}{\bibfnamefont{G.}~\bibnamefont{Mak}} \bibnamefont{and}
  \bibinfo{author}{\bibfnamefont{H.~M.} \bibnamefont{Vandriel}},
  \bibinfo{journal}{Phys. Rev. B} \textbf{\bibinfo{volume}{49}},
  \bibinfo{pages}{16817} (\bibinfo{year}{1994}).

\bibitem[{\citenamefont{Zhou et~al.}(1994)\citenamefont{Zhou, Vandriel, and
  Mak}}]{Zhou94}
\bibinfo{author}{\bibfnamefont{X.~Q.} \bibnamefont{Zhou}},
  \bibinfo{author}{\bibfnamefont{H.~M.} \bibnamefont{Vandriel}},
  \bibnamefont{and} \bibinfo{author}{\bibfnamefont{G.}~\bibnamefont{Mak}},
  \bibinfo{journal}{Phys. Rev. B} \textbf{\bibinfo{volume}{50}},
  \bibinfo{pages}{5226} (\bibinfo{year}{1994}).

\bibitem[{\citenamefont{Loren et~al.}(2011)\citenamefont{Loren, Rioux, Lange,
  Sipe, van Driel, and Smirl}}]{Loren11}
\bibinfo{author}{\bibfnamefont{E.~J.} \bibnamefont{Loren}},
  \bibinfo{author}{\bibfnamefont{J.}~\bibnamefont{Rioux}},
  \bibinfo{author}{\bibfnamefont{C.}~\bibnamefont{Lange}},
  \bibinfo{author}{\bibfnamefont{J.~E.} \bibnamefont{Sipe}},
  \bibinfo{author}{\bibfnamefont{H.~M.} \bibnamefont{van Driel}},
  \bibnamefont{and} \bibinfo{author}{\bibfnamefont{A.~L.} \bibnamefont{Smirl}},
  \bibinfo{journal}{Phys. Rev. B} \textbf{\bibinfo{volume}{84}},
  \bibinfo{pages}{214307} (\bibinfo{year}{2011}).

\bibitem[{\citenamefont{Uenoyama and Sham}(1990)}]{Uenoyama90}
\bibinfo{author}{\bibfnamefont{T.}~\bibnamefont{Uenoyama}} \bibnamefont{and}
  \bibinfo{author}{\bibfnamefont{L.~J.} \bibnamefont{Sham}},
  \bibinfo{journal}{Phys. Rev. Lett.} \textbf{\bibinfo{volume}{64}},
  \bibinfo{pages}{3070} (\bibinfo{year}{1990}).

\end{thebibliography}


\end{document}